\documentclass[showpacs,amsmath,amssymb,aps,pra,
longbibliography,
10pt,reprint,superscriptaddress]{revtex4-1}

\usepackage{bm}
\usepackage[breaklinks=true,colorlinks=true,linkcolor=blue,urlcolor=blue,citecolor=blue]{hyperref}
\usepackage{dcolumn}
\usepackage{amsmath,amssymb}
\usepackage{mathdots}
\usepackage{natbib}
\usepackage{soul,color}
\usepackage{graphicx}
\usepackage{epstopdf}
\usepackage[note-name]{notes2bib}
\usepackage{mathptmx}
\usepackage{stmaryrd}

\begin{document}

\title{\Large{\bf Fast Dynamics of Vortices in Superconductors}}

\author{Oleksandr\,V.~Dobrovolskiy}
    \email[Corresponding author: ]{oleksandr.dobrovolskiy@univie.ac.at}
    \affiliation{Superconductivity and Spintronics Laboratory,
                Nanomagnetism and Magnonics, Faculty of Physics,
                University of Vienna, W\"ahringer Str. 17,
                1090 Vienna, Austria}

\begin{abstract}
The last decade has been marked by great interest in the dynamics of vortices moving at high ($>$10\,km/s) velocities in superconductors. However, the flux-flow instability (FFI) prevents its exploration and sets practical limits for the use of vortices in various applications. Even so, FFI has turned into a valuable tool for studying the quasiparticle energy relaxation in superconductors, with the view of enhancement of the single-photon detection capability of micrometer-wide strips operated at large bias currents. In this context, the failure of ``global'' FFI models for materials with strong intrinsic and edge vortex pinning has urged the elaboration of ``local'' FFI models. This chapter outlines the recent advances in the research on FFI and highlights superconductors with perfect edges and weak volume pinning as prospective materials for studying fast vortex dynamics and nonequilibrium superconductivity. This manuscript is based on the author's version of the review published at \url{https://doi.org/10.1016/B978-0-323-90800-9.00015-9}.

\end{abstract}

\maketitle

\tableofcontents

\section{Introduction}
Fast vortex dynamics and related nonequilibrium phenomena are essential subjects of research in superconductivity\,\cite{Wor12prb,Pui12pcs,Che14apl,Gri15prb,Gur14prl,Buh15nac,Lar17pra,She17prb,Emb17nac,Mad18sca,Dob19pra,Bez19prb,Kog20prb,Dob20nac,Lya20nac,She20prb,Dob20pra,Kog21prb,Pat21prb,Kog22prb,Bud22pra}. High velocities of magnetic flux quanta (Abrikosov vortices or fluxons) attract great interest because of the fundamental questions regarding their stability as topological excitations of the superconducting order parameter and the ultimate speed limits for magnetic flux transport via vortices at intense transport currents. Furthermore, fast dynamics of fluxons determines the vortex-assisted mechanism of voltage response in superconducting single-photon detectors\,\cite{Vod17pra,Kor20pra} and makes accessible novel phenomena in superconductor-based systems, such as the generation of sound\,\cite{Ivl99prb,Bul05prb} and spin\,\cite{Bes14prb,Dob21arx} waves, as well as emerging functionalities for, e.g., microwave radiation/detection\,\cite{Bul06prl,Dob18nac,Loe19acs,Dob20cph}.

Probing the upper speed limits for vortices represents a valuable approach for studying the energy relaxation processes in superconductors\,\cite{Kni06prb,Leo11prb,Att12pcm,Leo20sst}. This method allows one to quantify the lifetimes of quasiparticles (unpaired electrons) when their energy distribution is out of equilibrium. The physical phenomenon underlaying this approach is termed  \emph{flux-flow instability} (FFI)\,\cite{Lar75etp,Lar86inb,Bez92pcs,Kun02prl}. In the current-voltage ($I$-$V$) curve of superconductors, FFI becomes apparent as a discontinuous jump to a highly resistive state at some instability current $I^\ast$ which corresponds to the instability voltage $V^\ast$, see Fig.\,\ref{f1}(a). Note that $I^\ast$ is usually smaller than the maximal current a superconductor can carry without dissipation -- the \emph{depairing current} $I_\mathrm{d}$, thereby setting the actual limit for the use of superconductors in applications. The instability voltage allows one to quantify the maximal vortex velocity (instability velocity) $v^\ast$ by using the standard relation $v^\ast = V^\ast/(BL)$, where $B$ is the magnitude of the applied magnetic field and $L$ the distance between the voltage leads. The understanding of the microscopic mechanisms determining the maximal vortex velocities and the behavior of $v^\ast(B)$ is an essential task in the FFI research, see Fig.\,\ref{f1}(b).

Reviews and discussions of FFI and nonlinear effects occurring during fast vortex motion are given in Refs.\,\cite{Att12pcm,Dea18smf,Hue19snm,Mis07prb,Kun03mpl,Vod20ltp}. This chapter presents (i) a summary of the different FFI models with emphasis on the links between them, (ii) an outline of vortex-pinning effects on FFI and the associated \emph{paradigm shift} from \emph{global}\,\cite{Lar75etp,Lar86inb,Bez92pcs} to \emph{local}\,\cite{Bez19prb,Vod19sst} FFI models, and (iii) some of the prospective research directions\,\cite{Mak22adm,Fom22apl,Dob22mmm}.

\section{Flux-flow instability models}
\subsection{Flux-flow instability mechanisms}
Theoretical foundations for understanding of the electronic nonequilibrium effects were laid by Eliashberg\,\cite{Eli70etp,Eli86inb}. Due to external energy input, the quasiparticle energy distribution is modified with respect to equilibrium, which affects the superconducting parameters such as critical temperature $T_\mathrm{c}$, critical current $I_\mathrm{c}$, and superfluid density, and can even result in their enhancement\,\cite{Wya66prl}. For instance, the quasiparticle energy distribution can be modified via irradiation with light, microwaves, phonons, injection of quasiparticles, etc.\,\cite{Gra81boo}.

Experimental studies of the photon absorption response and the effects of high-frequency irradiation of superconductors are usually performed at zero external magnetic field\,\cite{Nat12sst,Zol13ltp}. By contrast, nonequilibrium effects in the mixed state and the vortex motion under intense transport currents and FFI are studied in the presence of external magnetic fields\,\cite{Gra81boo}.

\begin{figure}[t!]
    \centering
    \includegraphics[width=1\linewidth]{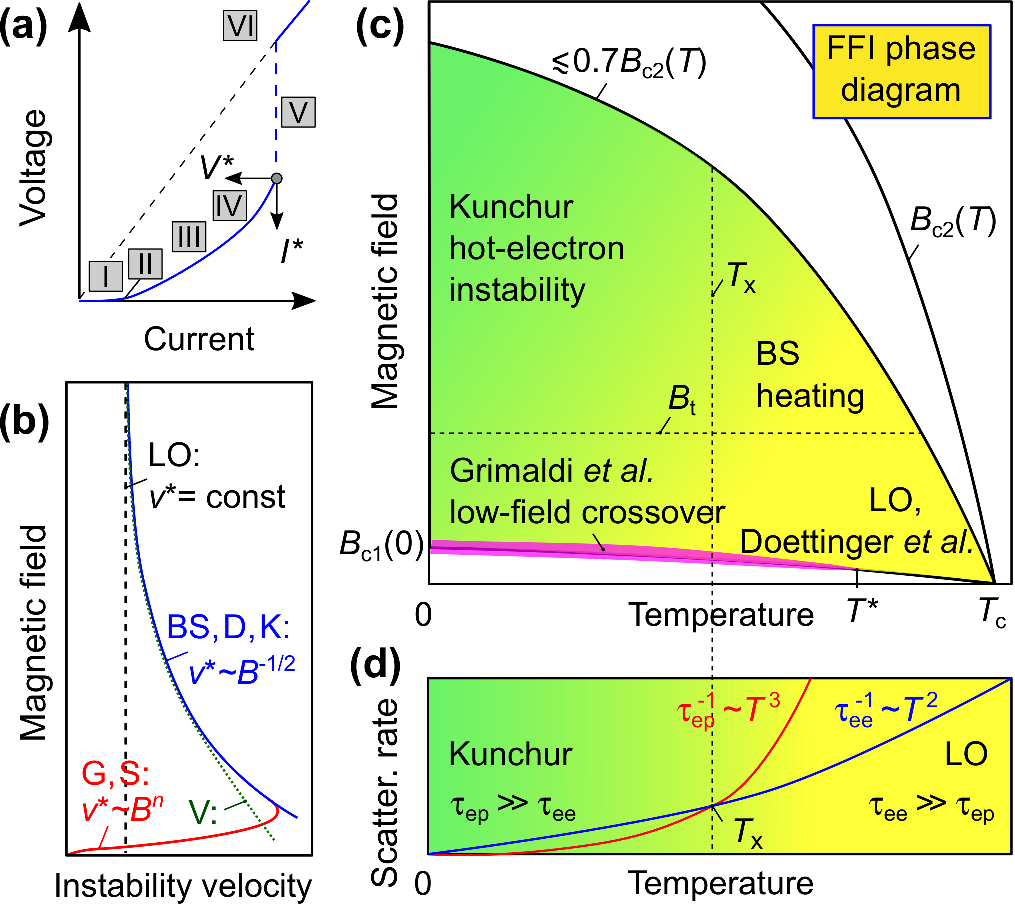}
    \caption{(a) Typical $I$-$V$ curve of a superconductor in the mixed state,
    with the indicated instability current $I^\ast$ and instability voltage $V^\ast$,
    and the regimes of pinned vortices (I), depinning transition (II), flux flow (III),
    nonlinear conductivity (IV), flux-flow instability (V), and highly-resistive state (VI).
             (b) Predictions of different FFI models for $v^\ast(B)$.
    LO: Larkin-Ovchinnikov\,\cite{Lar75etp},
    BS: Bezuglyj-Shklovskij\,\cite{Bez92pcs},
    D: Doettinger \emph{et al}\,\cite{Doe95pcs},
    K: Kunchur\,\cite{Kun01prl},
    G: Grimaldi \emph{et al}\,\cite{Gri10prb},
    S: Shklovskij\,\cite{Shk17pcs},
    V: Vodolazov\,\cite{Vod19sst}.
            (c) Different FFI regimes on the $B$-$T$ plane.
                $B_\mathrm{t}$: overheating field in the BS model\,\cite{Bez92pcs}.
            (d) Temperature dependences of the inelastic
       electron-electron and electron-photon scattering rates.
    $T_\mathrm{x}$: crossover temperature.}
    \label{f1}
\end{figure}

The FFI mechanisms are distinct at high ($T\approx T_\mathrm{c}$) and low ($T\ll T_\mathrm{c}$) temperatures, see Fig.\,\ref{f1}(c). Close to $T_\mathrm{c}$, FFI can be described by the Larkin-Ovchinnikov (LO) theory\,\cite{Lar75etp,Lar86inb}. In the LO theory, FFI is associated with a shrinking of the moving vortex and a reduction of the viscosity of the superconducting medium due to the diffusion of quasiparticles from the vortex core to the surroundings. Well below $T_\mathrm{c}$, the LO mechanism is ineffective since in this case superconducting energy gap $\Delta$ does not depend strongly on small changes of the electron distribution function.

At $T\ll T_\mathrm{c}$, FFI is described within the framework of the Kunchur (K) hot-electron model in which the electronic temperature $T_\mathrm{e}$ is elevated due to dissipation. The higher $T_\mathrm{e}$ stipulates the creation of additional quasiparticles, which diminishes $\Delta$. This leads to an expansion of the vortex core\,\cite{Kun01prl,Kun02prb}, which has the effect of reducing the viscous drag because of the softening of gradients of the vortex profile\,\cite{Kun02prl}.

The LO model\,\cite{Lar75etp} is justified when the inelastic electron-electron scattering time $\tau_\mathrm{ee}$ is much larger than the electron-photon scattering time $\tau_\mathrm{ep}$, $\tau_\mathrm{ee}\gg\tau_\mathrm{ep}$, which ensures a \emph{non-thermal} electron energy distribution. By contrast, $\tau_\mathrm{ee}\ll\tau_\mathrm{ep}$ in the K model implies a \emph{thermal-like} electron distribution so that the electronic ensemble exhibits a shift of the electronic temperature $T_\mathrm{e}$ with respect to the phonon temperature $T_\mathrm{p}$\,\cite{Kun01prl}.

\subsection{Quasiparticle scattering mechanisms}
Standard estimates\,\cite{Kap76prb,Abr88boo} for the scattering rates
$\tau_\mathrm{ee}^{-1} = k_\mathrm{B}T^2/r \hbar \varepsilon_\mathrm{F}$ and
$\tau_\mathrm{ep}^{-1} = k_\mathrm{B}T^3/r^3 \hbar T_\mathrm{D}^2$ [$\varepsilon_\mathrm{F}$: Fermi energy, $T_\mathrm{D}$: Debye temperature] give a crossover temperature $T_\mathrm{x} = r^2 k_\mathrm{B}T^2_\mathrm{D}/\varepsilon_\mathrm{F}$. Here, $r<1$ is the phonon reflection coefficient at the film-substrate interface, arising from their acoustic mismatch. Thus, one has $\tau_\mathrm{ee}>\tau_\mathrm{ep}$ above $T_\mathrm{x}$ which determines the crossover between the K and LO models, see Fig.\,\ref{f1}(d).

For classical superconductors, the electron-phonon interaction is usually dominant compared to the electron-electron interaction\,\cite{Aro78etp}. By contrast, the electron-electron interaction is more important in materials with a high $T_\mathrm{D}$ and in high-$T_\mathrm{c}$ superconductors\,\cite{Doe97prb}. For instance, $T_\mathrm{x}\simeq1$\,K for Nb makes accessible both K and LO regimes\,\cite{Leo11prb}. By contrast, $T_\mathrm{x}\simeq100$\,K for YBCO points to the dominating electron-electron scattering in the entire temperature range of the superconducting state\,\cite{Kni06prb}. Finally, $T_\mathrm{x}\simeq0.5$\,K for amorphous superconductors like MoGe\,\cite{Lia10ffi} suggests that the LO mechanism dominates in a broad range of temperatures of the superconducting state.

\subsection{Larkin-Ovchinnikov model of FFI}
\begin{figure}[t!]
    \centering
    \includegraphics[width=1\linewidth]{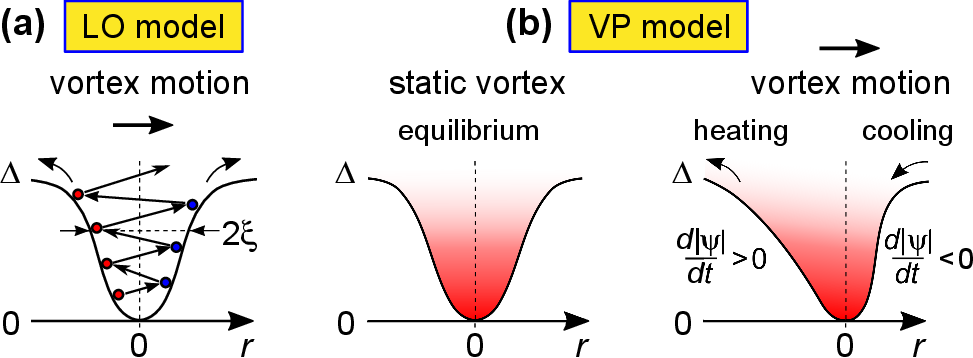}
    \caption{Schematics of the vortex core shapes in the LO\,\cite{Lar75etp,Kle85ltp}
    and VP\,\cite{Vod07prb} models assuming spatially uniform and nonuniform
    nonequilibrium quasiparticle distributions, respectively.
    (a) The electric field generated by fast vortex motion raises the energy
    of the quasiparticles trapped in the core. During this process the particle
    character alternates between electron-like and hole-like due to Andreev reflections.
    The vortex core shrinks in consequence of the quasiparticle escape.
    (b) Deformation of the vortex core due to vortex motion.
    The color depth indicates the density of the quasiparticles.
    In case the quasiparticle diffusion length is smaller than $\xi$,
    diffusion of the quasiparticles is not strong and locally there is
    an effective cooling and heating of the quasiparticles.
    }
    \label{f2}
\end{figure}

The key point of the LO theory\,\cite{Lar75etp,Lar86inb} is illustrated in Fig.\,\ref{f2}(a). At high vortex velocities, the quasiparticles within the vortex core are accelerated toward the core boundary, where they undergo Andreev reflection\,\cite{And64etp,Blo82prb,Oct83prb} and alternate between electrons and holes. As a result, their energy increases and they eventually escape from the vortex core. This leads to a shrinking of the vortex core and the vortex viscosity $\eta$ becomes a nonmonotonic function of the velocity $v$, namely $\eta(v) = \eta(0)/[1 + (v/v^\ast)^2]$\,\cite{Lar75etp,Lar86inb}, so that the viscous force $F_v = \eta(v) v$ has a maximum at $v^\ast$. At velocities exceeding $v^\ast$, $F_v$ decreases, leading to an even further increase of $v$, causing an instability in the vortex motion.

The nontrivial nature of the LO effect should be emphasized especially. If one uses the Bardeen-Stephen expression\,\cite{Bar65prv} for $\eta = \Phi_0^2/(2\pi\xi^2\rho_\mathrm{n})$ [$\Phi_0$: magnetic flux quantum, $\xi$: coherence length, $\rho_\mathrm{n}$: resistivity in the normal state], one finds that it actually \emph{increases} if the size of the vortex core ($\sim2\xi$) decreases. By contrast, in the LO model, the vortex core shrinks and this results in a \emph{decrease} of $\eta$. This contradiction can be explained by the failure of the Bardeen-Stephen model for the vortex as a normal cylinder with radius of the order of $\xi$ when the vortex is moving with a high enough velocity $v$.

The essential assumption of the LO theory is a \emph{spatially uniform} nonequilibrium quasiparticle distribution function $f(E)$ in the superconductor. Accordingly, while the vortex core shrinks as the vortex is accelerated to high velocities, its shape remains \emph{axially symmetric} in the LO model, see Fig.\,\ref{f2}(a). Moreover, FFI nucleates at all points of the superconductor at the same time and this leads to a \emph{field-independent} $v^\ast$ at which FFI occurs, see Fig.\,\ref{f1}(b). Thus, all refinements of the LO model dealing with the spatially uniform nucleation of FFI can be termed \emph{global} FFI models.

\subsection{Refinements of the Larkin-Ovchinnikov model}
Early experiments on low-$T_\mathrm{c}$ superconductors confirmed the LO prediction of FFI\,\cite{Mus80etp,Kle85ltp,Vol92fnt}. It was revealed that FFI occurs at magnetic fields $B\lesssim0.7B_\mathrm{c2}$\,\cite{Bab04prb}, where $B_\mathrm{c2}$ is the upper critical field of the superconductor. From the good quantitative agreement between theory and experiment in large magnetic fields, it was concluded\,\cite{Mus80etp,Kle85ltp} that the condition of spatially uniform $f(E)$ is well satisfied when the distance between vortices $a(B)\simeq\sqrt{\Phi_0/B}\ll L_\epsilon\equiv v^\ast\tau_\epsilon$ [$L_\epsilon$: quasiparticle diffusion length, $\tau_\epsilon$: quasiparticle energy relaxation time].

However, at $B\ll B_\mathrm{c2}$, $v^\ast(B)\sim \sqrt{1/B}$ was observed in many experiments\,\cite{Doe94prl,Doe95pcs,Lef99pcs}, see Fig.\,\ref{f1}(b). The observed dependence was explained as a consequence of the crossover from a uniform to a \emph{nonuniform} distribution of nonequilibrium quasiparticles when $L_\epsilon = \sqrt{D\tau_\epsilon}$ [$D$: electron diffusion coefficient] becomes comparable with the intervortex distance, $v^\ast\tau_\epsilon\simeq a(B)$. Doettinger \emph{et al} suggested\,\cite{Doe95pcs} that at smaller fields the system can be recovered to a spatially homogeneous state by allowing $v^\ast$ to grow accordingly to the increase of $a$ with decrease of the applied magnetic field, $a=(2\Phi_0/\sqrt{3}B)^{1/2}$. The LO expression was complemented\,\cite{Doe95pcs} with the term $a/\sqrt{D \tau_\epsilon}$, yielding
\begin{equation}
\label{e1}
    v^\ast = \left[\frac{(1-t)^{1/2}D[14\zeta(3)]^{1/2}}{\pi \tau_\epsilon}\right]^{1/2}
    \left(1 + \frac{a}{\sqrt{D \tau_\epsilon}}\right),
\end{equation}
where $t = T/T_\mathrm{c}$ and $\zeta(x)$ is the Riemann function. Note that a $v^\ast(B)\propto1/\sqrt{B}$ dependence appears in other FFI models as well.

In the original LO theory, the nonlinear flux-flow behavior is due only to the \emph{electric field-induced} change in $f(E)$ and not Joule heating. Bezuglyj and Shklovskij (BS) extended the LO theory in the thin-film configuration by taking into account \emph{Joule heating} effects\,\cite{Bez92pcs}. BS found that even if the Joule heating of the lattice is negligible, quasiparticles can still experience an overheating due to the finite heat removal rate of the power dissipated in the sample which can trigger FFI. Thus, in the BS model, $f(E)$ depends on the \emph{vortex density} and the \emph{rate of heat removal} from the film to the substrate. BS used the electronic temperature approximation, i.\,e. the assumption that the quasiparticles temperature $T_\mathrm{e}$ is uniform throughout the film thickness due to a high electron thermal conductivity. Thus, if the film thickness $d$ is much smaller than the electron-phonon scattering length (that usually occurs in thin films), the nonequilibrium phonons can escape from the film without re-absorption leading to the electron overheating\,\cite{Bez92pcs}.

BS introduced a transition magnetic field $B_\mathrm{t}$, see Fig.\,\ref{f1}(c), which depends on the heat transfer coefficient between the film and the substrate\,\cite{Bez92pcs}. At fields $B<B_\mathrm{t}$ the hypothesis of a uniform quasiparticles distribution in the LO model remains justified. By contrast, at $B\gg B_\mathrm{t}$ dissipation during the flux flow raises the electronic temperature $T_\mathrm{e}$ and the onset of FFI is governed by thermal effects. The BS model yields $v^\ast(B)\propto B^{-1/2}$, see Fig.\,\ref{f1}(b). Note just the same dependence $v^\ast(B)\propto B^{-1/2}$ is found in the K model\,\cite{Kun01prl} at $T\ll T_\mathrm{c}$ in which thermal effects diminish the superconducting order parameter and lead to an expansion of the vortex cores. However, in the low-temperature regime, the quasiparticles heating reduces $B_\mathrm{c2}$ to $B_\mathrm{c2}(T_\mathrm{e})<B_\mathrm{c2}$.

\subsection{Quasiparticle relaxation: Dirty vs clean regimes}
The schematics of the vortex core in Fig.\,\ref{f2}(a) allows one to consider distinct quasiparticle energy relaxation process for dirty and clean superconductors\,\cite{Per05prb}. When the mean electron free path $l$ is shorter than the coherence length (dirty limit, $l< \xi$), the energy relation occurs via the \emph{scattering process inside the vortex core}. In the opposite case (clean limit, $l>\xi$) the quasiparticles can relax their energy by \emph{recombination outside the vortex core} only after covering a distance within the core equivalent to several $\xi$.

In the microscopic picture where the quasiparticle energy rise comes from Andreev reflections\,\cite{And64etp,Blo82prb,Oct83prb} at the vortex core boundary, see Fig.\,\ref{f2}(a), an increase of the quasiparticles energy occurs well below $v^\ast$ in clean samples so that $v^\ast$ values in the clean limit are noticeably smaller than in the dirty limit\,\cite{Per05prb}. Indeed, in the dirty limit, more energy is needed to induce the depletion of quasiparticles of the core. The shrinking of the vortex core occurs at higher energy and $\tau_\epsilon$ is substantially \emph{temperature independent}. In the clean limit, the quasiparticle distribution function is influenced by the recombination of quasiparticles to Cooper pairs. This means that in the clean limit $\tau_\epsilon$ is strongly \emph{temperature dependent} since the quasiparticles recombination mechanism obeys an \emph{exponential law}\,\cite{Doe97prb,Per05prb,Lia10ffi}.

\subsection{High-$T_\mathrm{c}$ superconductors and ultraclean regime}
FFI was also extensively investigated in high-temperature superconductors (HTSs). These studies were performed for epitaxial\,\cite{Doe94prl,Doe95pcs,Xia96prb,Che14apl}, vicinal\,\cite{Pui12pcs} and He ion-irradiated\,\cite{Xia96ltp} YBa$_2$Cu$_3$O$_{7-\delta}$ films, MgB$_2$\,\cite{Kun03prb},
Bi$_2$Sr$_2$CaCu$_2$O$_{8+\delta}$\,\cite{Xia98prb2}, La$_{1.85}$Sr$_{0.15}$CuO$_{4-x}$\,\cite{Doe97prb}, Nd$_{2-x}$Ce$_x$CuO$_y$\,\cite{Sto98prl,Hue99prb,Sto99prb} films and YBa$_2$Cu$_3$O$_{7-\delta}$ nanowires\,\cite{Lya18prb,Rou18mat}. The distinct effects of intrinsic (nonequilibrium) and extrinsic (thermal runaway) mechanisms driving FFI were discussed for YBa$_2$Cu$_3$O$_{7-\delta}$ films\,\cite{Maz08prb} and for iron based superconductors\,\cite{Leo16prb}. Signatures of FFI in the anisotropic FeSeTe have allowed for its treatment halfway between low-$T_\mathrm{c}$ and high-$T_\mathrm{c}$ superconductors.

In the dirty limit, the energy distribution of the quasiparticles in the vortex core is strongly smeared out due to their high scattering rate. The separation between the quasiparticles energy levels and between the Fermi energy $\varepsilon_\mathrm{F}$ and the lowest bound state in the vortex core is proportional to $\xi^{-2}$\,\cite{Hue19snm}. With an estimate $\xi\lesssim 10$\,nm for disordered low-temperature superconductors\,\cite{Por19acs,Kor20pra,Cor19nal}, the core can be treated as an energy continuum of quasiparticles. By contrast, the extremely small $\xi\lesssim1$\,nm in the HTS cuprates makes the \emph{electron quantum structure} of the vortex core essential. For its experimental observation, the energy smearing $\delta\varepsilon = \hbar/\tau$ due to the mean electronic scattering time $\tau$ must be sufficiently small, $\delta\varepsilon\ll \Delta^2/\varepsilon_\mathrm{F}$, corresponding to the ultraclean superconducting limit. Such a regime has been experimentally realized in Refs.\,\cite{Sto98prl,Hue99prb,Sto99prb}. Proceeding from an isolated vortex to the vortex lattice, the discrete energy levels within the vortex core interact between neighboring vortices and broaden into \emph{minibands}\,\cite{Hue19snm}. In this case, vortex motion at high velocities is accompanied by emerging phenomena, such as Bloch oscillations of the quasiparticles, negative differential resistance, and multi-step instabilities\,\cite{Hue19snm,Sto98prl,Hue99prb,Sto99prb}.

\section{Pinning effects on the flux-flow instability}
\subsection{Effects of uncorrelated disorder on FFI}

The LO model considered a moving vortex lattice in an infinite superconductor in the Wigner-Seitz approximation and ignored collective effects related to vortex pinning and lattice transformations. The assumption of no pinning corresponds to a $\delta$-functional velocity distribution in the vortex ensemble and a free flux flow with zero critical current\,\cite{Lia10prb}. In this case, an increase of current produces a shift of the velocity distribution to higher mean values $\langle v \rangle$ until the FFI occurs at $v^\ast$ for all vortices \emph{simultaneously}. Note, it is the \emph{mean value} $\langle v \rangle$ which is deduced from the dc voltage measured experimentally.

\begin{figure}[t!]
    \centering
    \includegraphics[width=0.9\linewidth]{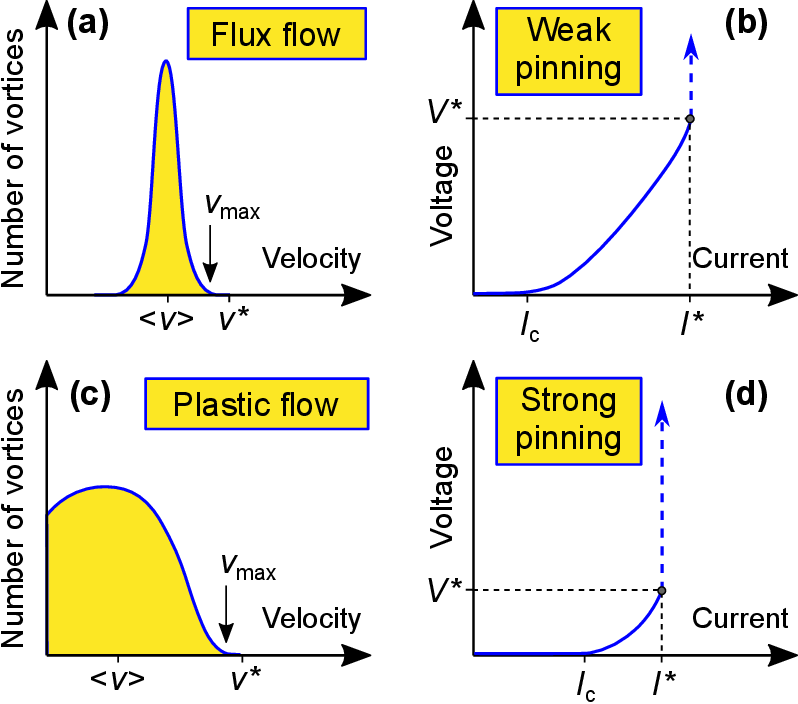}
    \caption{Schematics for different dynamic regimes after\,\cite{Sil12njp}.
            Velocity distribution for the case of flux flow (a) and plastic flow (c).
            Expected $I$-$V$ curves for weak (b) and strong (d) pinning strengths.
            The critical current density $I_\mathrm{c}$ indicates the onset of dissipation.
            The instability current $I^\ast$ and voltage $V^\ast$ indicate where FFI takes place.}
    \label{f3}
\end{figure}

In disordered systems, a rather narrow distribution of vortex velocities, see Fig.\,\ref{f3}(a), can be realized in a \emph{weak-pinning} regime or at high enough fields and currents when the intervortex interaction dominates the vortex-pinning interaction\,\cite{Bla94rmp}. In particular, an increase of the current $I$ above $I_\mathrm{c}$ produces a shift of the velocity distribution toward higher mean values $\langle v \rangle$ and progressively \emph{narrows} the distribution peak in consequence of the enhanced long-range order in the moving vortex lattice\,\cite{Kos94prl}. The FFI jump occurs at the end of the quasilinear flux-flow regime, see Fig.\,\ref{f3}(b). The commonly occurring vortex dynamics regimes are indicated in Fig.\,\ref{f1}(a).

The introduction of \emph{strong pinning} by \emph{uncorrelated disorder} substantially modifies the previous picture\,\cite{Ruc00prb,Leo10pcs,Sil12njp,Shk17prb}. As pointed out by Silhanek \emph{et al}\,\cite{Sil12njp}, on the one hand, the presence of disorder broadens the velocity distribution and extends the nonlinear regime for currents above the critical value $I_\mathrm{c}$\,\cite{Gri09prb}. This is schematically shown in Fig.\,\ref{f3}(c) and (d). On the other hand, strong pinning increases $I_\mathrm{c}$ and causes FFI to occur within the depinning-mediated nonlinear section of the $I$-$V$ curve. Importantly, the broadening of the velocity distribution implies a \emph{sizable separation between} $\langle v \rangle$ and the maximal attainable vortex velocity $v^\ast$. This means that it is the vortices which move faster than other vortices in the ensemble which actually trigger FFI. Once FFI is triggered in the regions of faster-moving vortices it is then spread over the entire volume of the superconductor.

Thus, in this plastic vortex flow regime, the system reaches the instability point at a \emph{lower} vortex velocity than expected, practically when $v_\mathrm{max}\simeq v^\ast$. In this case, the LO assumption of homogeneous quasiparticle distribution in the whole sample has to be dropped and instead homogeneous quasiparticle distribution only along the paths where the vortices move has to be considered\,\cite{Sil12njp}. As the magnetic field increases, the distance between the vortices decreases and the vortex-vortex interaction increases, which in turn effectively diminishes the role of pinning. This narrows the vortex velocity distribution so that $v^\ast$ increases. With a further increase of the magnetic field, the assumption of homogeneous quasiparticle distribution becomes justified again, and the LO scenario is retained. This gives rise to qualitatively different dependences of $v^\ast(B)$, see Fig.\,\ref{f1}(b), and enables controlling the flux-flow dissipation via vortex pinning engineering in superconductors\,\cite{Ruc00prb,Gri12apl}.

\subsection{Low-field crossover in the instability velocity}
A crossover to a different behavior of $v^\ast(B)$ at magnetic fields $B < B_\mathrm{cr1}$ was reported by Grimaldi \emph{et al}\,\cite{Gri10prb,Gri09pcm} for Nb films with moderately strong intrinsic pinning. This field $B_\mathrm{cr1}$ corresponds to a crossover from the regime where $v^\ast$ \emph{increases} with increasing magnetic field to the regime $v^\ast \propto B^{-1/2}$, see Fig.\,\ref{f1}(b). Magneto-optical imaging of the flux penetration into the films at $B\ll B_\mathrm{cr1}$ revealed that the flux does not penetrate with a smooth advancing front, but instead as a series of irregularly shaped protrusion. This causes the formation of an inhomogeneous magnetic field distribution and induces preferential stationary channels for the moving vortices with a total effective width $l$\,\cite{Gri10prb}.

An inspection of the films by field emission scanning electron microscopy revealed that these channels are grouped together in clusters along the strip\,\cite{Gri10prb}. This means that the voltages measured across the distance $L$ between the voltage leads are actually produced across an \emph{effective distance} $l < L$, so that in the low-field range $B<B_\mathrm{cr1}$, the flux-flow channels fill only parts of the whole strip length.

By contrast, when the flux penetration is smooth, above some threshold temperature $T^\ast$ ($\simeq0.75T_\mathrm{c}$ for the films studied in\,\cite{Gri10prb}), the pinning strength decreases so the vortex flow becomes uniform as well. In this way, below $T^\ast$, the decreasing low-field dependence $v^\ast(B)$ is preserved, whereas above $T^\ast$, the crossover field disappears, leading to a monotonous decrease of $v^\ast(B)\propto B^{-1/2}$, see Fig.\,\ref{f1}(b) and (c). Herewith, $B_\mathrm{cr1}$ corresponds to a crossover from the channel-like vortex motion to a uniform flux-flow state with increase of the magnetic field. Accordingly, at $T<T^\ast$, the flux-flow resistance $\rho_\mathrm{ff}$ assumes lower values than the Bardeen-Stephen prediction while the expected linear $\rho_\mathrm{ff}(B)\propto B$ is retained at $T>T^\ast$\,\cite{Gri10prb}.

\subsection{Pinning effects on self-heating and FFI}
The effects of pinning and self-heating on FFI in thin films were considered theoretically by Shklovskij \emph{et al} for the cases $T\ll T_\mathrm{c}$\,\cite{Shk17pcs} and $T\approx T_\mathrm{c}$\,\cite{Shk17prb}. The problem was considered in a single-vortex approximation for the nonlinear $I$-$V$ curves derived for a saw-tooth\,\cite{Shk06prb} and cosine\,\cite{Shk08prb} pinning potentials.

In the low-temperature regime, within the framework of the K model\,\cite{Kun02prl}, the presence of pinning has been revealed\,\cite{Shk17pcs} to not modify the magnetic field dependences of the electric field $E^\ast(B)$, current density $j^\ast(B)$ and resistivity $\rho^\ast$ at the instability point, which remain monotonic. By contrast, the derivative of the instability velocity $d v^\ast/d B$ may \emph{change its sign twice}, as observed in experiments\,\cite{Leo10pcs,Gri12apl,Sil12njp,Gri09pcm,Gri10prb,Gri11snm,Dob17sst,Leo20ltp}. In particular, experiments on weak-pinning $5\,\mu$m-wide Mo$_3$Ge thin films\,\cite{Leo20ltp} revealed that the observed nonmonotonic $v^\ast(B)$ dependence can be described well with the theoretical dependences of Ref.\,\cite{Shk17pcs}. For wider films, that is in the absence of the surface pinning caused by the mesoscopic geometry, the experimentally observed\,\cite{Leo20ltp} dependence $v^\ast(B)\propto B^{-1/2}$ fits the K model\,\cite{Kun01prl}.

In the high-temperature regime, within the framework of the LO theory\,\cite{Lar75etp} and its generalization by BS\,\cite{Bez92pcs}, the increase of the pinning strength does not change the field dependences of the electric field strength $E^\ast(B)$ and the current density $j^\ast(B)$ qualitatively. At a fixed magnetic field value, $E^\ast$ and $v^\ast$ \emph{decrease} while $j^\ast$ \emph{increases} with increasing magnetic field. In this way, the theoretical analysis of Refs.\,\cite{Shk17pcs,Shk17prb} corroborates that the \emph{presence of pinning} leads to a \emph{reduction} of $v^\ast$, with exception of highly-correlated, periodic pinning-induced regimes in the vortex dynamics as outlined next.

\subsection{Vortex lattice commensurability effects on FFI}
In the presence of a periodic pinning, the spatial commensurability between the vortex lattice with the pinning landscape leads to magnetoresistance minima and critical current maxima\,\cite{Luq07prb,Vil03sci,Dob12njp,Sol14prb,Dob20pra,Jaq02apl,Dob17sst,Dob18nac}. In this regime, $v^\ast$ attains a \emph{maximum} when the spacing between the rows of vortices driven across an array of nanogrooves is commensurate with the distance between them\,\cite{Dob17sst}. The location of this matching peak can be tuned through the nanostructure period variation.

The considered case can be realized for nanopatterned superconductors at the fundamental matching field when vortex dynamics is in a coherent regime, i.\,e. the entire vortex ensemble behaves as a vortex crystal\,\cite{Luq07prb,Vil03sci,Sol14prb,Dob20pra}. At the same time, the background pinning due to undesired random disorder must be weak to ensure the long-range order in the vortex lattice in the vicinity of the depinning transition. This condition can be realized, e.g., in weak-pinning amorphous Mo$_3$Ge\,\cite{Gri15prb}, Nb$_7$Ge$_3$\,\cite{Bab04prb}, MoSi\,\cite{Bud22pra}, nanocrystalline Nb-C\,\cite{Dob20nac}, as well as Al\,\cite{Sil12njp} and epitaxial Nb films in the clean superconducting limit\,\cite{Dob12tsf}.

\subsection{Fast dynamics of guided magnetic flux quanta}
\begin{figure}[t!]
    \centering
    \includegraphics[width=0.9\linewidth]{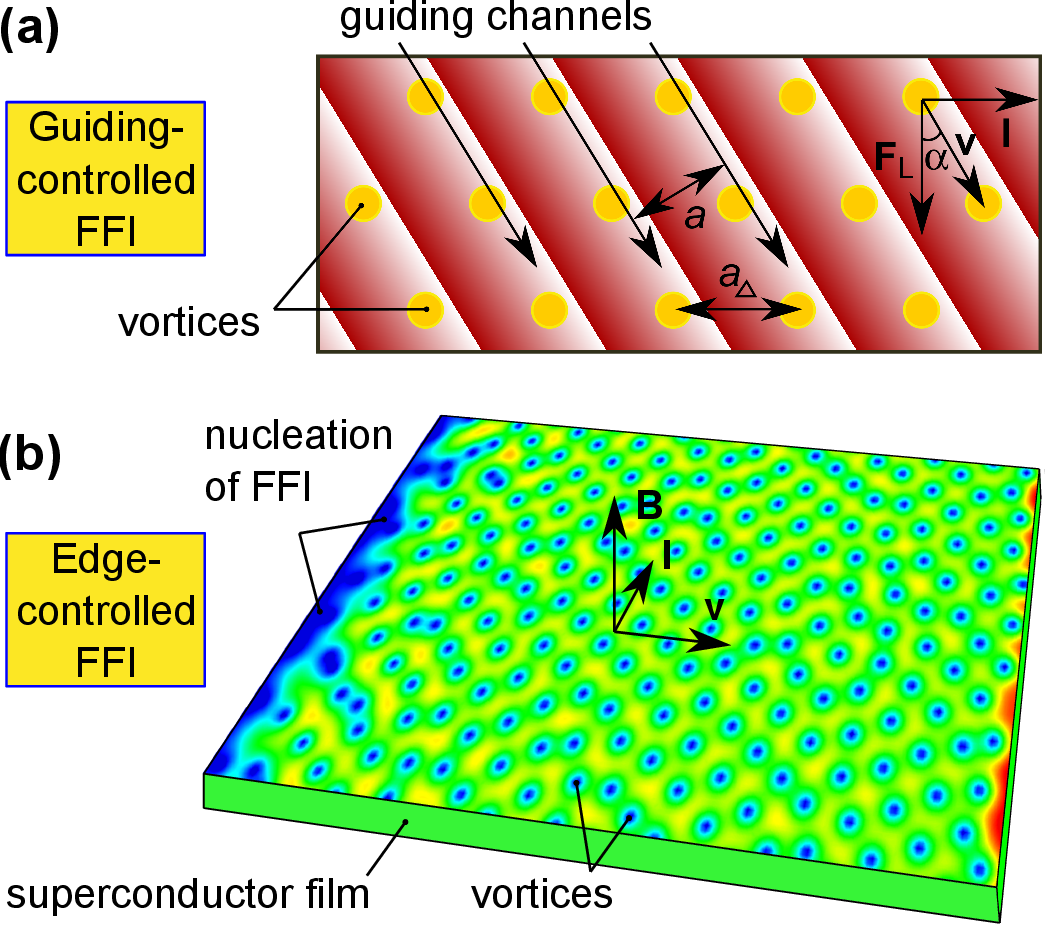}
    \caption{(a) Schematic of a vortex lattice which is commensurate with
    a periodic washboard pinning landscape. When the vortices are guided at some
    tilt angle $\alpha$ with respect to the pinning channels,
    the long-range order in the vortex lattice is enhanced\,\cite{Dob19pra,Dob20pra}.
    (b) The combination of a weak intrinsic pinning, close-to-depairing critical current and fast relaxation of nonequilibrium allows for ultra-high vortex velocities\,\cite{Dob20nac}. The nucleation of FFI occurs at the film edge of the film and is decisively controlled by its quality\,\cite{Vod19sst,Bud22pra}.
    }
    \label{f4}
\end{figure}

An interesting possibility to enhance the long-range order in the moving vortex lattice is offered by the vortex guiding effect\,\cite{Nie69jap,Shk06prb,Dob10sst}. This effect consists in the noncollinearity of the velocity of vortices with the driving force exerted on them by the transport current. In the regime of guiding under the action of the driving Lorentz-type force $\mathbf{F}_\mathrm{L}$ directed at a small tilt angle $\alpha$ with respect to linearly extended ``pinning sites'', see Fig.\,\ref{f4}(a), all vortices move along the pinning channels and the distribution of their velocities is close to the $\delta$-functional shape. The small angle $\alpha$ ensures that the normal component of the driving force $F_\mathrm{L}\sin\alpha$ is not large enough to let vortices overcome the potential troughs, while the tangential component $F_\mathrm{L}\cos\alpha$ is large enough to drive the vortices at a few km/s velocities\,\cite{Dob19pra}.

The presence of the vortex velocity component \emph{along} the sample length in the guiding regime leads to the appearance of a \emph{transverse} voltage. This voltage vanishes upon a transition from the vortex guiding regime to a regime in which vortices start overcoming the pinning potential barriers\,\cite{Shk06prb,Rei08prb}. By guiding magnetic flux quanta at a tilt angle of $15^\circ$ with respect to Co nanostripe arrays deposited on top of Nb films, a fivefold enhancement of $v^\ast$ (up to 5\,km/s at low magnetic fields) was observed experimentally\,\cite{Dob19pra}. In a different experiment, $I$-$V$ measurements on tracks perpendicular to the vicinal ($8^\circ$) step direction of YBa$_2$Cu$_3$O$_{7-\delta}$ films (current crossed between $ab$ planes) provided evidence for the sliding motion along the $ab$ planes (vortex channeling) at velocities exceeding $8$\,km/s\,\cite{Pui12pcs}.

\section{From global to local instability models}

\subsection{Local FFI model for thin films near $T_\mathrm{c}$}
Distinct from the nonlinear conductivity regime considered by LO, FFI jumps were also observed in linear $I$-$V$ sections\,\cite{Vol92fnt,Gri11snm}. A broad distribution of vortex velocities caused by the presence of regions with different pinning strengths is a likely cause for such a regime. Namely, on average, an essentially larger number of slowly moving vortices can make a larger contribution to the measured voltage as compared to the contribution of a much smaller number of faster moving vortices, allowing the $I$-$V$ curve to maintain a linear shape up to the instability point. The respective \emph{local} FFI theory was developed by Bezuglyj \emph{et al}\,\cite{Bez19prb}, based on the assumption that FFI occurs upon reaching $I^\ast$ not in the entire superconducting film but in a narrow strip with weak pinning across the film width in which vortices move much faster than outside it.

As a consequence of the overheating of the local areas with faster moving vortices, \emph{normal domains} can be formed \emph{across} the superconducting film. Once the current density exceeds some threshold current density (determining the equilibrium of the nonisothermal normal/superconducting boundary)\,\cite{Gur84spu,Bez84ltp} these domains begin to grow and the whole sample transits to the normal state\,\cite{Bez19prb}. In this case, the standard relation $v^\ast = V^\ast/(B L)$ can no longer yield the instability velocity quantitatively. Nevertheless, the functional relations between the LO instability parameters in the case of a local FFI remain the same as in the case of a global FFI, but require \emph{renormalization} of the sample length\,\cite{Bez19prb}.

\subsection{Rearrangement of vortices in the TDGL model}
The LO model assumes that the vortex lattice does not exhibit any structural changes upon the transition of the superconductor to a state with a resistance close to the normal value. However, experiments on low- and high-temperature superconductors suggested that some kind of phase transition may occur in the fast-moving vortex lattice and regions with fast and slow vortex motions may appear in the sample\,\cite{Gri15prb,Ada15prb}.

An alternative view of the moving vortex at high velocities, in which it loses the shape of a rigid cylinder, was used by Vodolazov and Peeters (VP)\,\cite{Vod07prb}. VP solved \emph{numerically} the generalized time-dependent Ginzburg-Landau (TDGL) equations\,\cite{Kra78prl,Wat81ltp} which take into account the spatially \emph{nonuniform} distribution of quasiparticles as the longitudinal (odd in energy $E$) part $f_\mathrm{L}(E) = f(-E) - f(E)$ of the nonequilibrium $f(E)$ distribution is localized only in the region where the time derivative $\partial|\psi|/\partial t$ is finite, that is only near the moving vortex core, see Fig.\,\ref{f2}(b).

Indeed, the motion of a vortex implies a suppression of the order parameter $\psi$ in front of the vortex and its recovery behind it. If the vortex velocity is large enough, $v\simeq\xi/\tau_\epsilon$, the number of quasiparticles in front of the vortex becomes smaller than the equilibrium value and it becomes larger behind the vortex due to the finite relaxation time of $f(E)$. Effectively, this can be viewed as \emph{cooling} of the quasiparticles in front of the vortex and \emph{heating} behind the vortex. Since the relaxation time of $\psi$ depends on temperature as $\tau_{|\psi|}\sim1/(T_\mathrm{c}-T)$, the healing time of the order parameter behind the vortex is long while the time of the order parameter suppression in front of the vortex is short. This leads to an \emph{elongated shape} of the vortex core with a point where $|\psi|=0$ displaced in the direction of the vortex motion.

The elongated shape of the core of a fast-moving vortex is in line with the field distribution around it as derived \emph{analytically} based on the time-dependent London (TDL) equation (which ignores the vortex core)\,\cite{Kog20prb}. The flux quantum of a moving vortex is redistributed: The back-side part of the flux is enhanced, whereas the in-front part is depleted\,\cite{Kog21prb}. Distortions of the field distribution of single moving vortices lead to distorted intervortex interactions and therefore to a change in the vortex lattice structure\,\cite{Kog18prb}. At large velocities, the moving vortex lattice adopts the structure with one of the lattice vectors along the velocity, the same result as obtained within TDGL\,\cite{Lid04prb,Vod07prb}.

\subsection{Edge-controlled FFI in weak-pinning regime}
In K\,\cite{Kun02prl,Kni06prb,Shk17pcs} and BS\,\cite{Bez92pcs,Shk17prb} models the temperature of the electrons $T_\mathrm{e}$ was calculated from the balance between the spatially- and time-averaged Joule dissipation and the heat removal to the substrate, which ignores the discrete nature of the moving vortices. Vodolazov (V)\,\cite{Vod19sst} solved the TDGL equation in conjunction with the heat conductance equation for $T_\mathrm{e}$ and the energy balance equation to find the phonon temperature $T_\mathrm{p}$, which were derived from the \emph{kinetic equations} for the electron and phonon distribution functions\,\cite{Vod17pra}.

In V model, in the low-resistive state, there is a temperature gradient across the width of the strip with maximal local temperature near the edge where vortices enter the sample\,\cite{Vod19sst}. The higher temperature at the edge is caused by the larger current density in the near-edge area due to the presence of the \emph{edge barrier} for vortex entry and, hence, the locally larger Joule dissipation. With increase of the current, there is a series of \emph{transformations} of the moving vortex lattice in V model.

At $I\lesssim I^\ast$, localized areas with strongly suppressed superconductivity and closely spaced vortices appear near the hottest edge (left edge in Fig.\,\ref{f4}(b)). Upon reaching $I^\ast$, these areas begin to grow in the direction of the opposite edge and form highly resistive Josephson SNS-like links -- \emph{vortex rivers} -- along which the vortices move. Such vortex rivers were observed experimentally by scanning local probes\,\cite{Sil10prl,Emb17nac}. Due to the increasing dissipation, vortex rivers evolve into normal domains which then expand along the strip\,\cite{Dob20nac}.

Distinct from the modified LO model\,\cite{Doe95pcs}, in the V model $v^\ast \sim B^{-1/2}$ only at comparatively large magnetic fields (see Fig.\,\ref{f1}(b)), when the intervortex distance at $I\simeq I_\mathrm{c}$ and $I\simeq  I^{\ast}$ is almost the same despite the transformations in the moving vortex lattice. By contrast, at relatively small magnetic fields, $a$ in the vortex rows is smaller than $(2\Phi_0/B\sqrt{3})^{1/2}$ at $I\simeq I^{\ast}$ and the number of vortices $n$ is smaller than follows from $n \Phi_0 = B S$ [$S$: sample area]. This leads to a weaker dependence $v^{\ast}(B)$ than $v^{\ast} \sim B^{-1/2}$ following from ``global'' instability models. This behavior is confirmed experimentally\,\cite{Dob20nac}.

\subsection{Edge barrier effects on the fluxon speed limits}
The prediction\,\cite{Vod17pra} and experimental confirmation\,\cite{Kor18pra,Kor20pra,Cha20apl,Chi20apl} of the single-photon detection capability of micrometer-wide strips has turned FFI into a widely used method for judging whether a superconducting material could be potentially suitable for single-photon detection\,\cite{Lin13prb,Cap17apl,Dob20nac,Cir21prm}. This assessment is based on the deduction of the quasiparticle relaxation times from FFI. However, while many dirty superconductors (NbN\,\cite{Kor20nph}, MoSi\,\cite{Kor20pra}, NbRe\,\cite{Cir20apl}, NbReN\,\cite{Cir21prm} etc.) possess good single-photon detection capability, $v^\ast$ and $\tau_\epsilon$ deduced from FFI are not always consistent with the fast relaxation processes implied by photon-counting experiments\,\cite{Bud22pra}.

The edge quality turns out to be decisive for the nucleation of FFI and, hence, quantification of the relaxation times from the $I$-$V$ measurements\,\cite{Bud22pra}. By investigating FFI in wide MoSi strips differing only by the edge quality, a factor of 3 larger critical currents $I_\mathrm{c}$, a factor of 20 higher maximal vortex velocities of 20\,km/s, and a factor of 20 shorter $\tau_\epsilon$ have been revealed for MoSi films with perfect edges. Thus, for the deduction of the intrinsic $\tau_\epsilon$ of the material from the $I$-$V$ curves, utmost care should be taken regarding the edge and sample quality, and such a deduction is justified only if the field dependence of $I_\mathrm{c}$ points to the dominating edge pinning of vortices\,\cite{Bud22pra}.

\subsection{Edge defects as gates for fast-moving fluxons}
The requirement of close-to-depairing critical currents $I_\mathrm{c}$ in superconductor strips is related to blocking of the penetration of vortices via the strip edges and knowledge of the effects of various edge defects on the penetration and patterns of Abrikosov vortices\,\cite{Gla86ltp,Ala01pcs,Vod03pcs,Vod15sst,Siv18ltp}. Once the Lorentz force exerted on a vortex by the transport current exceeds the force of attraction of the vortex to the sample edge, the edge barrier is suppressed. This suppression can be local in the case of a local increase of the current density -- \emph{current-crowding} effect\,\cite{Ada13apl}, and it can be realized, e. g., in strips with an edge defect\,\cite{Dob20nac,Bud22pra}. In this case, the defect acts as a gate\,\cite{Ala01pcs} for vortices entering into the superconductor strip and crossing it under the competing action of the Lorentz and vortex-vortex interaction forces.

If the size of the defect is much larger than $\xi$, the defect can serve as a nucleation point for several vortex chains\,\cite{Emb17nac,Bud22pra}. Such chains form a \emph{vortex jet} with the apex at the defect and expanding due to the repulsion of vortices as they move to the opposite film edge. If the defect size is $\simeq\xi$, the vortices enter into the strip consequentially. However, in the presence of fluctuations and inhomogeneities in the strip, the vortex chain evolves into a diverging jet because of the intervortex repulsion.

Vortices penetrating into a superconducting Pb film at rates of tens of GHz and moving with velocities of up to tens of km/s were observed by scanning SQUID-on-tip (SOT) microscopy\,\cite{Emb17nac}. Though these vortices move faster than the perpendicular current superflow which drives them, no excessive changes in the structure of the Abrikosov vortex core has been revealed even at velocities of the order of 10\,km/s\,\cite{Emb17nac}. Because of the large current density gradient across the constriction with narrowing, a vortex chain splits up into a fan-like pattern of slipstreaming vortices. The TDGL modeling suggested that the slipstreaming vortices could further evolve to Josephson-like vortices moving with velocities as high as $\sim100$\,km/s\,\cite{Emb17nac}. Spatial modulation of the order parameter between these vortices is rather weak and the channel behaves effectively as a self-induced Josephson junction, which appears without materials weak links. Such flux channels in thin films can be interpreted as phase-slip lines\,\cite{Siv03prl}.

Vortex jet shapes have recently been analyzed based on the dynamic equation\,\cite{Bez22prb}. For wide strips (width $w$ much larger than the penetration depth $\lambda$), the analytically calculated vortex jet shapes reproduce qualitatively the experiment\,\cite{Emb17nac}. For narrow strips ($\xi\ll w <\lambda$), the discrete TDGL approach reveals dynamic transitions from a vortex chain to a vortex jet and, at $I\lesssim I^\ast \simeq I_\mathrm{d}$, from the vortex jet to a vortex river\,\cite{Bez22prb}.

Finally, it is interesting to note that vortex jets can also be steered by using an inhomogeneous magnetic field. For instance, in the recently studied 3D superconductor open nanotubes (tubes with a slit) with an azimuthal transport current and an external magnetic field applied perpendicular to the tube axis, the TDGL modeling has revealed a suppression of the vortex chain/vortex jet transition with increase of the current and/or field and the vortex jets were constraint to the tube areas where the normal component of the applied magnetic field was close to maximum\,\cite{Bog23arx}.

\section{Fast vortex dynamics at a microwave ac stimulus}

\subsection{Microwave stimulation of superconductivity}
Microwave irradiation is being used to control the quantum properties of different systems, ranging from supercurrents in superconductors to mechanical oscillators\,\cite{Ber10prl,Mci12nan,Pal13nat,Cle20nph,Bar22nph}. Nowadays, using nonequilibrium pumping for cooling is a hot topic\,\cite{Bha14nac,Jos14nac,For17nan,Mai20nac,Sch20nan,Uro20sci}. In 1966, microwave stimulated superconductivity (MSSC) was discovered in superconducting bridges\,\cite{Wya66prl} and later confirmed for different type I superconducting systems\,\cite{Day67prv,Pal79prb,Tol83etp,Hes93prb,Pal80prl,Bec13prl,Vis14prl}. It was   later found that also acoustic waves\,\cite{Tre75prl} and tunneling injection currents may, under certain conditions, cause an increase of the gap in the excitation  spectrum, the order parameter, the critical current and the  critical temperature $T_\mathrm{c}$\,\cite{Gra81boo}.

The phenomenon of MSSC was explained by Eliashberg (E)\,\cite{Eli70etp} as a consequence of an irradiation-induced redistribution of quasiparticles away from the gap edge. Namely, in the BCS theory, there is a fundamental connection between the gap value $\Delta$ and the occupation of the quasiparticle states\,\cite{Bar57prv,Tin04boo}. When the temperature is increased from zero, more and more quasiparticles are excited blocking states which were previously available for Cooper pair formation. Ultimately this process leads to the disappearance of superconductivity at $T_\mathrm{c}$. Extraction of quasiparticles leads to enhancement of superconductivity but is elusive to realize.  However, there is another possibility. Electron states near the Fermi surface are more important to Cooper pair formation than states further away from it. If a quasiparticle in a state near $\varepsilon_\mathrm{F}$ is removed to  a state with higher energy, the energy gap increases, although the total number of quasiparticles remains constant. This is the essential basis of the E mechanism. In a microwave field, the quasiparticles are ``pumped up'', away from the gap. Even if the average energy of the quasiparticles increases, the gap is enhanced. Recent reviews on MSSC are given in Refs.\,\cite{Kla20aop,Tik20aop}.

\subsection{Condensate and quasiparticles at dc and ac}

The vortex state is characterized by a spatially modulated superconducting order parameter which vanishes in the vortex cores and attains a maximal value between them. Accordingly, a type II superconductor can be regarded as a continuum medium consisting of bunches of quasiparticles in the vortex cores surrounded by a bath of the superconducting condensate formed by the superfluid of Cooper pairs. So far there is no available theory addressing the complex interaction of the superimposed dc and microwave currents with the quasiparticles and the condensate at the same time. However, the essential ingredients of this interaction have been studied separately.

\begin{figure}[t!]
    \centering
    \includegraphics[width=0.75\linewidth]{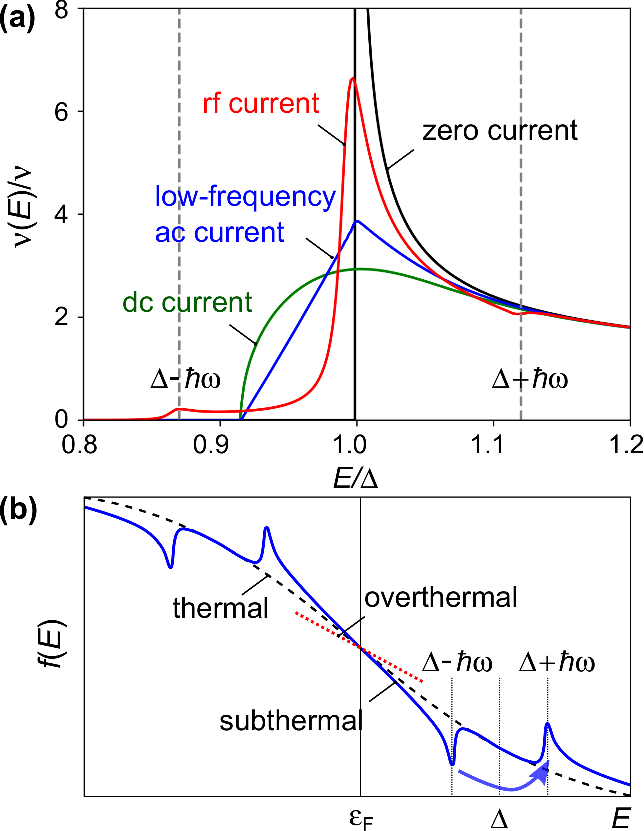}
    \caption{(a) Schematics of the modification of the electronic distribution function $f(E)$
    in the presence of a microwave excitation after\,\cite{Tik18prb}. The redistribution of the quasiparticles
    between the $\Delta - \hbar\omega$ and $\Delta + \hbar\omega$ states leads to a higher inclination of $f(E)$
    at the Fermi energy level $\varepsilon_\mathrm{F}$, which can be viewed as an effective ``cooling''.
    (b) Time-averaged density of states in a dirty superconductor at low temperature ($T\ll T_\mathrm{c}$)
    under different biasing after\,\cite{Sem16prl,Tik20aop}.
    The value of the current or its amplitude is $0.25I_\mathrm{c}$.}
    \label{f5}
\end{figure}

The effect of a dc current on the superconducting condensate is well known\,\cite{Ant03prl}. With an increase of a dc current, the absolute value of the order parameter is decreasing and the peak in the density of states at the edge of the superconducting gap is smeared, see Fig.\,\ref{f5}(a). This is because of gaining a finite momentum by the Cooper pairs that form a coherent excited state that plays a central role in the explanation of the gauge invariance of the Meissner effect\,\cite{And58prv}. A general theory of depairing by a microwave field has been formulated recently\,\cite{Sem16prl}, under the condition of rather small ac field amplitudes. It has been shown that the ground state of a superconductor is altered qualitatively in analogy to the depairing due to a dc current. However, in contrast to dc depairing, the density of states acquires steps at multiples of the microwave photon energy and shows an exponential-like tail in the subgap regime\,\cite{Sem16prl}. Additionally, depending on temperature, one can consider two regimes in which the response of a superconductor is dominated either by the response of the superfluid ($T\ll T_\mathrm{c}$) or by the quasiparticles ($T\approx T_\mathrm{c}$). It is also known that at $T\approx T_\mathrm{c}$, microwave radiation can be absorbed by quasiparticles, leading to a non-equilibrium distribution over the energies\,\cite{Eli70etp}, see Fig.\,\ref{f5}(b).

In general, the response of the condensate to an external microwave field becomes apparent via a change of the kinetic impedance (imaginary part of the complex resistivity) while the quasiparticles give rise to the microwave loss (real part of the complex resistivity). The known effects of stimulation of superconductivity by a microwave stimulus in the absence of an external magnetic field include an enhancement of the Ginzburg-Landau depairing current (in narrow channels) implying a transition to a resistive state due to the formation of phase-slip centers\,\cite{Zol13ltp} or the Aslamazov-Lempitskii maximum current (in wide films) at which the vortex structure induced by the self-field evolves into the first phase-slip line\,\cite{Asl82etp}. In the presence of an external magnetic field inducing vortices in the superconductor, the vortex-induced microwave losses dominate the response of the superconductor\,\cite{Pom08prb}.

\subsection{Microwave stimulation in the vortex state}
From Fig.\,\ref{f5}(a) it follows that a dc current leads to a smearing of the BCS peak at the gap edge. Further, the presence of vortices leads to a strongly spatially non-uniform order parameter (with normal-state regions inside the vortex cores). Moreover, the motion of current-driven vortices leads to dissipation\,\cite{Bra95rpp,Pom08prb}. All together, these circumstances diminish the stimulation effect of the microwave excitation. Nonetheless, in the regime of low vortex densities and for superconductors with rather slow relaxation of nonequilibrium quasiparticles, the stimulation of superconductivity under a microwave ac stimulus can be observed in the presence of vortices\,\cite{Lar15nsr,Dob19rrl}.

Note, the dynamics of vortices under a microwave ac stimulus differs qualitatively from the dc-driven vortex motion. Namely, while a moderately strong dc transport current (i.\,e. when the current-induced Lorentz-type force exceeds the pinning force\,\cite{Bra95rpp}) causes a \emph{translational} motion of vortices, a microwave ac current \emph{shakes} the vortices. Accordingly, in the presence of pinning sites, one can distinguish between the low-frequency and high-frequency regimes in the vortex dynamics. Namely, when the ac half-period is long enough, a vortex can move over \emph{many} pinning sites, while in the high-frequency regime a vortex is shaken in vicinity of \emph{one} pinning site. The crossover between the two regimes occurs at the so-called \emph{depinning frequency} $f_\mathrm{d}$\,\cite{Git66prl,Cof91prl,Pom08prb,Shk08prb,Shk14pcm,Dob15met}. In the low-frequency regime, the pinning forces dominate and the vortex response is weakly dissipative. By contrast, in the high-frequency regime, the oscillating vortices cease to feel the detail of the pinning potential, the frictional forces prevail, and the response is strongly dissipative. Note, this consideration is justified for short vortices in thin films and it is inapplicable to thick films where long vortices can get elastically distorted extending over many pinning sites\,\cite{Pat21prb}.

The depinning frequency depends on temperature $f_\mathrm{d}(T,H,I) \simeq f_\mathrm{d}(0,H,I)[1 - (T/T_\mathrm{c})^4]$\,\cite{Zai03prb,Ali20sst}, magnetic field $f_\mathrm{d}(T,H,I) \simeq f_\mathrm{d}(T,0,I)[1 - (H/H_\mathrm{c2})^2]$ \cite{Jan06prb,Ali20sst} and the dc current as $f_\mathrm{d}(T,H,I) \simeq f_\mathrm{d}(T,H,0)[1- (I/I_\mathrm{d})^{m}]^{n}$ with the exponents $m$ between $3/2$ and $4$ and $n$ between $1/4$ and $2/3$, depending on the details of the pinning potential\,\cite{Dob17nsr}. At relatively low excitation power levels, for a commensurate vortex lattice in a periodic pinning landscape, ac losses due to ac-driven vortex oscillations are minimal because of the enhanced efficiency of the pinning\,\cite{Dob15apl,Dob20pra}. By contrast, a qualitatively different behavior is observed when the microwave power increases beyond a certain, frequency-dependent power level: the microwave losses are maximal for pinned vortices, and they are minimal for depinned vortices\,\cite{Lar15nsr,Dob19rrl}. In this nonlinear regime, the vortices act as ``relaxators'' rather than rigid ``oscillators''. Since quasiparticles escape from the vortex cores, their escape can be seen as an effective decrease of the amplitude of the vortex oscillations with increase of the ac frequency.

\subsection{Braking of instabilities by a microwave stimulus}
\begin{figure}[t!]
    \centering
    \includegraphics[width=1\linewidth]{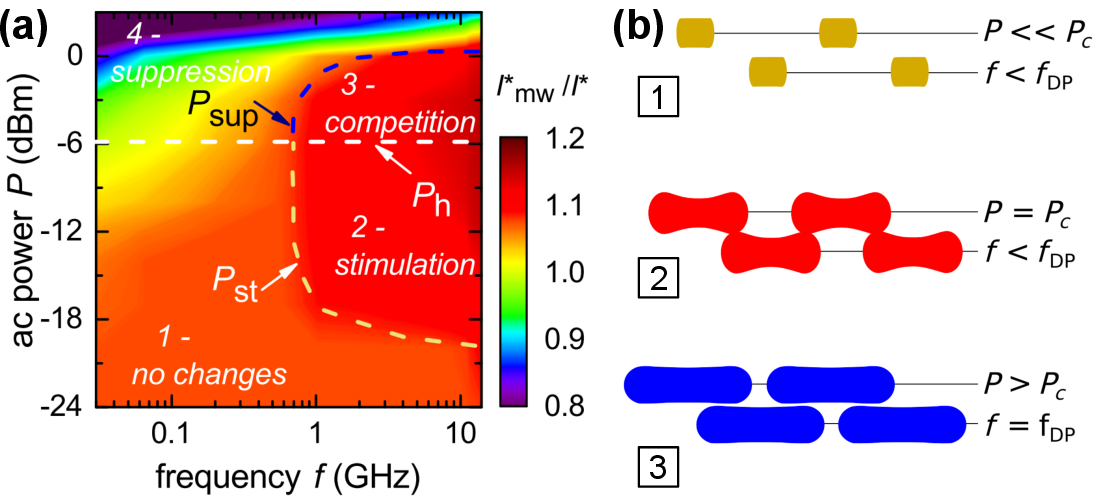}
    \caption{(a) Evolution of the instability current $I^\ast_\mathrm{mw}$ for a Nb film at $T = 0.998 T_\mathrm{c}$ and $H = 10$\,mT
            in a broad range of frequencies and power levels of the ac stimulus.
            $I^\ast$: instability current value without microwave stimulus.
            Adapted under the terms of the CC-BY Creative Commons Attribution 4.0 International license\,\cite{Dob20cph}.
            Copyright 2020, The Authors, published by Springer Nature.
       (b) The sketch illustrates the overlap of the regions where vortices move under a microwave ac field of different powers and frequencies.
            As the vortices move, their radii change. This could lead the heated zones to overlap and produce avalanches.
            Adapted with permission\,\cite{Lar17pra}. Copyright 2017, American Physical Society.
    }
    \label{f6}
\end{figure}

The FFI onset can be advanced or delayed in the presence of superimposed dc and ac current drives\,\cite{Che14apl,Dob20cph}. Namely, in the presence of a dc current, an abrupt transition to a high-loss state occurs at lower microwave power levels\,\cite{Che14apl}. The addition of an ac stimulus on top of a dc current allows for tuning the FFI onset\,\cite{Dob20cph}, see Fig.\,\ref{f6}(a), depending on the ac power $P$ and frequency $f$. Thus, when the ac power exceeds some threshold value $P_\mathrm{st}$, at high enough (GHz) ac frequencies the quasilinear flux-flow regime can be extended up to a higher current $I^\ast_\mathrm{mw}$, whereas the FFI occurs at a smaller $I^\ast_\mathrm{mw}$ in the presence of a low-frequency ac current\,\cite{Dob20cph}. With a further increase of the ac power, the heating will dominate the cooling effect and the FFI will occur at a smaller $I^\ast_\mathrm{mw}$. This behavior can be explained, qualitatively, based on a model of ``breathing mobile hot spots'', implying a competition of heating and cooling of quasiparticles along the trajectories of moving fluxons whose core sizes vary in time. This breathing mode appears owing to the variation of the vortex core sizes in time due to the periodically modulated quasiparticle escape from the cores. Accordingly, the strong oscillations of vortices lead to the formation of ``clouds'' of quasiparticles around the vortex cores, whose relaxation takes place in a larger volume as compared to the non-excited case. Thus, dissipation is decreasing due to the combined effects of the redistribution of quasiparticles away from the gap edge via the E mechanism and the relaxation of the quasiparticles in larger volumes around the vortex cores. A theory of MSSC at high vortex velocities is yet to be elaborated.

Instabilities can also occur in superconductors when the flux is getting redistributed upon temperature or magnetic field variation\,\cite{Col21sst,Ara05prl,Alt04rmp,Min81rmp}. Herewith, vortex penetration may occur either smoothly or in sudden bursts, giving rise to \emph{thermomagnetic instabilities}. In the case of smooth penetration, the flux density inside the superconductor changes gradually, being described by a \emph{critical state model}\,\cite{Bea62prl,Kim63prv2,Zel94prl}. However, when the penetration is abrupt, narrow branches of flux, in the form of dendrites, invade the sample. The critical state model is no longer appropriate to describe this case.

In a typical critical state type of flux distribution, millions of vortices self-organize, under the competing vortex-vortex repulsion, Lorentz and pinning forces. This delicate balance results in metastable states between which sudden flux redistributions occur. These \emph{flux avalanches} occur typically during a slow ramping of the applied magnetic field, and at temperatures below a certain fraction of $T_\mathrm{c}$. The flux patterns are never reproduced when experiments are repeated, thus ruling out possible explanations based on material defects guiding the flux motion\,\cite{Col21sst}. The thermomagnetic instability or flux jumping arises because of two fundamental reasons\,\cite{Den06prl}: (i) motion of magnetic flux releases energy, and hence increases the local temperature; (ii) the temperature rise decreases flux pinning, and hence facilitates the flux motion. This positive feedback can result in thermal runaways and global flux redistributions jeopardizing superconducting devices.

According to the model based on thermomagnetic instabilities\,\cite{Rak04prb}, an avalanche is triggered by a thermal fluctuation (hot spot), which facilitates more flux motion toward the hot place, with a subsequent heat release. Under quasiequilibrium conditions, the avalanches can also be triggered by microwave pulses\,\cite{Sol13jap} or ac signals at near resonant frequencies of superconducting cavities\,\cite{Ghi07jap}. Under a broadband microwave field sweep, flux avalanches are triggered at different depinning frequencies due the distribution in strength of individual vortex pinning sites\,\cite{Awa11prb}.

At $f \lessgtr f_\mathrm{d}$, flux avalanches exhibit a thermally driven behavior since higher temperatures facilitate the entrance of magnetic flux and enhance the vortex mobility. Accordingly, the critical microwave power $P_\mathrm{c}$ for triggering an avalanche monotonously decreases with approaching $T_\mathrm{c}$. However, at $f \simeq f_\mathrm{d}$, the behavior of $P_\mathrm{c}(T)$ inverts, demonstrating a thermally driven avalanche \emph{inhibition}\,\cite{Lar17pra}. This effect can be understood as a consequence of the dependence of the vortex core size on $P$ and $f$. Namely, at $f \lessgtr f_\mathrm{d}$, the LO effect manifests itself only in the regions of maximal velocity (i.e., close to displacement minima), see panel 1 in Fig.\,\ref{f6}(b). On the other hand, at $f\simeq f_\mathrm{d}$, the vortices become more mobile. As a result, the LO reduction of the vortex core manifests itself during the whole displacement cycle, see panel 2 in Fig.\,\ref{f6}(b). Below $f_\mathrm{d}$ and at low $P$, local ``normal'' regions created by the driven vortices do not overlap and are not sufficient to trigger the avalanche process.

Increasing the microwave power below $f_\mathrm{d}$ enhances the absolute changes in the vortex core size, inducing the overlap between different extended core areas and triggering the avalanche at $P \simeq P_\mathrm{c}$, see panel 3 in Fig.\,\ref{f6}(b). This process is thermally activated since the vortex diameter strongly increases as $T \rightarrow T_\mathrm{c}$. However, at $f\simeq f_\mathrm{d}$ and in the temperature region where vortices become depinned, the LO effect reduces the vortex core ending up in only a small variation in the vortex core size during periodically driven motion. This reduces the probability of the overlap between normal regions and, therefore, the avalanche is triggered at higher microwave powers for a given temperature\,\cite{Lar17pra}.

\section{Emerging research directions}
\subsection{Superconductivity in 3D nanoarchitectures}
3D nanoarchitectures have become of increasing importance across various domains of science and technology\,\cite{Fom21boo,Mak22adm,Fom18inb}, including magnetism\,\cite{Fer17nac,Skj20nrp}, photonics\,\cite{Fre10afm}, magnonics\,\cite{Gub19boo} and plasmonics\,\cite{Win17ami}. From the applications viewpoint, the extension of nanoscale superconductors into the third dimension allows for the full-vector-field sensing in quantum interferometry\,\cite{Mar18nal}, noise-equivalent power reduction in bolometry\,\cite{Loe19acs} and reduction of footprints of fluxonic devices\,\cite{Fou11sst}. Extending quasi-1D or -2D superconductor manifolds into the third dimension allows for their thermal decoupling from the substrate and the development of multi-terminal devices and circuits with complex interconnectivity. Herewith, curved geometries bring about the smoothness of conjunctions, which allows for avoiding undesired weak links at sharp turns\,\cite{Por19acs} and minimizing current-crowding effects\,\cite{Cle11prb}.

The complex interplay between Meissner screening currents, Abrikosov vortices and slips of the phase of the superconducting order parameter in curved 3D nanoarchitectures gives rise to topological modes which do not occur in planar 2D films\,\cite{Fom22apl}. Nowadays, topological transitions between the different regimes in the vortex and phase-slip dynamics in curved 3D nanoarchitectures are a subject of extensive investigations\,\cite{Fom12nal,Cor19nal,Fom22nsr,Bog23arx}. The stability of these regimes under intense dc transport currents and high ac frequencies requires experimental and theoretical exploration.

\subsection{Cryogenic magnonics and magnon fluxonics}
The interplay between the fundamental excitations in superconducting and magnetic systems has recently given birth to the research domains of cryogenic magnonics and magnon fluxonics\,\cite{Gol18afm,Gol19pra,Dob19nph,Dob22mmm,Kna23jap}. Magnonics is one of the most rapidly growing research fields in modern magnetism\,\cite{Bar21pcms}. It is concerned with the dynamics of spin waves (and their quanta -- magnons), which are precessional excitations of ordered spins in magnetic materials, and their use for wave-based information processing\,\cite{Chu22tom}. In this regard, Meissner screening currents offer interesting possibilities to control the magnetization dynamics in superconductor-based hybrids at low temperatures\,\cite{Gol18afm,Gol19pra}. Furthermore, the periodic modulation of the local magnetic field emanating from the vortex cores induces Bloch-like bandgaps in the magnon frequency spectrum while a current-driven vortex lattice acts as a moving Bragg grating\,\cite{Bar96prl}, leading to Doppler shifts of the magnon bandgap frequencies\,\cite{Dob19nph,Dob22mmm}.

The fast motion of the vortex lattice opens new prospects for the magnon-fluxon interaction: the phenomenon of Cherenkov radiation of magnons\,\cite{Bul05prl,She11prl,Bes14prb}, when the vortex lattice velocity is equal to the phase velocity of the spin wave. The experimentally observed magnon emission is unidirectional (spin wave propagates in the direction of motion of the vortex lattice), monochromatic (magnon wavelength is equal to the vortex lattice parameter), and coherent (spin-wave phase is self-locked due to quantum interference with eddy currents in the superconductor)\,\cite{Dob21arx}. The magnon radiation is accompanied by a magnon Shapiro step in the $I$-$V$ curve of the superconductor, reduces the dissipation\,\cite{Bul05prl,She11prl,Bes14prb}, and inhibits the FFI onset\,\cite{Dob21arx}. Thus, the generation of collective modes by a fast-moving vortex lattice can be viewed as an interesting approach for the inhibition of FFI and studies of fast vortex dynamics.

\subsection{Fast dynamics of few and single fluxons}
According to Fig.\,\ref{f1}(b), the highest vortex velocities are expected for the low/zero-$B$ regime and the key challenge consists in determining the number of fluxons moving in the superconductor. Namely, the actual number, $n_\mathrm{v}$, of fluxons in the superconductor can be smaller (e.g. for strong edge barriers\,\cite{Dob20nac}) or larger (e.g. in the case of edge defects\,\cite{Bez22prb}) than follows from the simple estimate $n\Phi_0 = BS$\,\cite{Bev23pra}. Here, $n=1,2,..$ is the expected vortex number, $\Phi_0$ the magnetic flux quantum, and $S$ the area of the sample surface.

An approach for the quantitative determination of $n_\mathrm{v}$ and $v^\ast$ has recently been demonstrated experimentally\,\cite{Bev23pra,Bev23rrl,Ust23etp}. The idea is based on the Aslamazov and Larkin (AL) prediction\,\cite{Asl75etp} of kinks in the $I$-$V$ curves of wide and short superconducting constrictions when the number of vortices crossing the constriction is increased by one. Such conditions were realized for a-few-micrometer-wide MoSi\,\cite{Bev23pra,Bev23rrl} and MoN\,\cite{Ust23etp} thin strips in which slits, milled by a focused ion beam\,\cite{Hof23arx}, locally suppress the edge barrier and allow for a controllable vortex entry. The edge-barrier asymmetry gives rise to negative differential resistance\,\cite{Ust22etp2} and superconducting diode effects\,\cite{Ust22etp}, opening up prospects for studies of vortex guiding\,\cite{Nie69jap,Shk06prb,Dob10sst} and ratchet\,\cite{Cer13njp,Shk11prb,Dob20pra} effects in the few- and single-fluxon regimes.

\section{Conclusion and outlook}
Summing up, the exploration of ultrafast vortex dynamics in superconductors has recently turned into a major research avenue. These studies are motivated by the enhancement of current-carrying capacity of superconductors, reduction of microwave loss therein, and improvement of superconducting single-photon detectors. From the viewpoint of basic research, the fast-moving ensemble of quantized magnetic flux lines entails emerging nonequilibrium phenomena and complex interactions, and it allows for the generation of collective modes in superconductor-based systems. It is anticipated that in the years to come, research along these lines will fuel the domain of vortex matter under nonequilibrium conditions and will lead to the development of novel applications.

\vspace{5mm}
\textbf{Acknowledgments}. This work is based on the research funded by the German Research Foundation (DFG) through Grant Nos. DO1511/2-1, DO1511/2-4, DO1511/3-1 and the Austrian Science Fund (FWF) through Grant Nos. I 4889 (CurviMag) and I 6079 (FluMag). Further, support by the European Cooperation in Science and Technology via COST Actions CA19108 (HiSCALE), CA19140 (FIT4NANO), and CA21144 (SUPERQUMAP) is acknowledged.

\vspace{5mm}
This manuscript is based on the author-prepared version of the review submitted to Elsevier
and published in Encyclopedia of Condensed Matter Physics, vol.\,\textbf{2} (2024) 735-754\,\cite{Dob23inb}.

\clearpage

\let\oldaddcontentsline\addcontentsline
\renewcommand{\addcontentsline}[3]{}


%

\end{document}